\documentstyle[epsfig,twoside,fleqn,espcrc2]{article}

\newcommand{\eq}{\begin{equation}}
\newcommand{\en}{\end{equation}}
\newcommand{\eqa}{\begin{eqnarray}}
\newcommand{\ena}{\end{eqnarray}}

\newcommand{\NP}[1]{Nucl.\ Phys.\ {\bf #1}}
\newcommand{\PL}[1]{Phys.\ Lett.\ {\bf #1}}
\newcommand{\PR}[1]{Phys.\ Rev.\ {\bf #1}}

\newcommand{\AmS}{{\protect\the\textfont2
  A\kern-.1667em\lower.5ex\hbox{M}\kern-.125emS}}

\title{Spectrum of the gauge Ising model in three dimensions.}

\author{
M. Caselle\thanks{Speaker at the conference}
\address{
Dip. di Fisica Teorica, Universit\`a di Torino,
Via P. Giuria 1, 10125 Torino, Italy}, 
M.Hasenbusch\address{Humboldt Universit\"at zu Berlin, Institut f\"ur Physik
Invalidenstr. 110, D-10099 Berlin, Germany}
and P.Provero$^{\rm a}$}%

\begin{document}

\begin{abstract}
We present a high precision Monte Carlo study of the  spectrum of the	
$Z_2$ gauge theory in three dimensions in the confining phase.
Using state of the art Monte Carlo techniques we are able to accurately 
determine up to three masses in a single channel. We compare our results
with the SU(2) spectrum and with the prediction of the Isgur-Paton model.
Our data strongly support the conjecture that the
glueball spectrum is described by some type of flux tube model.
We also compare the spectrum with some recent results for the correlation 
length in the 3d spin Ising model. This analysis
 sheds light on some nontrivial features of the duality 
transformation. 
\end{abstract}

\maketitle

\section{Introduction}
The infrared regime of
Lattice Gauge Theories (LGT) in the confining phase  displays a large 
degree of universality.
The main evidences  in favour of  this universality 
are given by the behaviour of the Wilson loop and  of the 
adimensional ratio $T_c^2/\sigma$ (where $T_c$ denotes the
 deconfinement temperature) which are roughly independent
from the choice of the gauge group, and show a rather 
simple dependence on the number of space-time 
dimensions. 
Both these behaviours are commonly understood as 
consequences of the fact  that the relevant degrees of freedom in the  
confining regime are string-like excitations. 
The phenomenological models which try to keep into account this string-like
picture are usually known as ``flux-tube'' models and turn out to give a very 
 good description of the Wilson loop behaviour 
(see for instance~\cite{ip} and
 references therein). 
The reason of this success is that in the interquark potential we
have a natural scale, the string tension, which allows to define in a
rather precise way a large distance (``infrared'') regime in which the 
adiabatic approximation for the string-like excitations can be trusted. 
This regime can be reached by considering interquark distances large in 
units of the string tension.

Besides the Wilson loop and the deconfinement temperature, another important set
of physical observables in LGT is represented by the glueball spectrum.
In this case it is less obvious that  a string like description could be used to
understand the data.
However, a  string-inspired model exists also for the glueball 
spectrum: the Isgur Paton model~\cite{ip} (IP in the following).

For this reason it would be very interesting to test if the same universality
(which, as mentioned above, should manifest itself as a substantial independence
from the choice of the gauge group) displayed by the Wilson loops also 
holds for the glueball spectrum. 
In this case we do not have 
the equivalent of the interquark distance, {\em i.e.}
a parameter which can be adjusted to select the infrared region:
the role of large Wilson loops is played by the higher states of the 
spectrum which, being localized in larger space regions, are expected to 
show more clearly a string-like behavior.
A major problem in this respect is the lack of
precise and reliable data for these higher states. 
An obvious proposal to overcome this problem is to begin with 
the (2+1)
dimensional case, for which some relevant simplifications occur in the spectrum
and a much higher precision can be achieved in the Montecarlo
simulations. 

Following this suggestion we have studied in~\cite{acch} the glueball spectrum
in the case of the (2+1) Ising gauge
model
obtaining  a high precision estimate of the first 11 states of the spectrum.
These can be compared with some results for  the (2+1) dimensional SU(2) model 
 obtained with Montecarlo simulations~\cite{tep} and with
 variational techniques~\cite{arisue}. 
The comparison between the SU(2) and Ising spectra shows that, 
not only the pattern of the states is the same in the two models, but also 
the values of the masses (except for the lowest state) are in remarkable
agreement.
This is a strong evidence in favour of the above mentioned universality, and
suggests that the higher states of the glueball spectrum of any LGT, 
(as it happens for the behaviour of large enough Wilson loops)
 can be predicted by some relatively simple flux-tube inspired model.
We shall give in the next section few general information on the model and on
the algorithm that we used (we refer to~\cite{acch} for further details). 
Sect 3.
is devoted to a discussion of the glueball spectrum and to a comparison with the
SU(2) results and with the IP predictions. Finally in the last section we shall
make some concluding remark on the duality transformation of the glueball
spectrum.

\section{The Ising gauge model}
The Ising gauge model is defined by the action
\eq
S_{gauge} =  - \beta \;\; \sum_{n,\mu<\nu} \;\; g_{n;\mu\nu}
\label{Sgauge}
\en
where 
$g_{n;\mu\nu}$ are the 
plaquette variables, defined in terms of the link 
fields $g_{n;\mu} \in {-1,1}$ as:
\eq
g_{n;\mu\nu}=g_{n;\mu} \; g_{n + \mu;\nu} \;
g_{n + \nu;\mu} \; g_{n;\nu}~~~.
\label{plaq}
\en
 where $ n \equiv (\vec x,t)$
denotes the space-time position of the link and $\mu$ its direction.
 For the Ising model, as for the SU(2) model, we cannot define a charge
conjugation operator. The glueball states are thus labelled only by their 
angular momentum $J$ and by their parity eigenvalue $P=\pm$. 
The standard notation is $J^{P}$. An important simplification due to the fact
that we are working in (2+1) dimensions is that in this case
all the states with 
angular momentum different from zero are degenerate in parity. Namely $J^+$ and
$J^-$ (with $J\neq 0$) must have the same mass. This result holds in the
continuum limit. The lattice discretization breaks this degeneration, since in 
this case the symmetry group is only the $D_4$ (dihedral) group. In particular
it can be shown that the degeneration still holds on the lattice 
for all the odd $J$ states, and is lifted for all the even $J$ states
(see~\cite{acch} for details). An important test of the whole analysis is to see
if the degeneration of the even part of the spectrum is recovered in the 
continuum limit. There are two other important consequences 
of the fact the symmetry group is reduced to $D_4$. The first one is that only 
operators with angular momentum $J~(mod(4))$ can be 
observed and the second is that all the odd $J$ states are grouped together 
in the same irreducible representation of $D_4$. 
This means that we cannot distinguish among 
them on the basis of the lattice symmetries.
Hence in the
following we shall  denote the states belonging
to this channel as $J=1/3$ states.

The simulations were performed 
 with a local demon-algorithm in Multi-spin-coding technique. 
Variational 
 techniques were used to 
 obtain accurate results 
 for the higher states of the spectrum in any given channel.
 We studied five different values of $\beta$
 in the scaling region, and carefully tested that all the states of 
the spectrum followed the expected scaling behaviour. As mentioned above  
 we obtained  reliable estimates for 11 masses.

\section{Results and comparison with SU(2) and with the IP model.}
A first unambiguous result of the simulation is that the parity degeneration is
indeed recovered in the continuum limit also for the even $J$ sector. Taking
into account this degeneration we end up with 8 independent states in the
continuum limit. Their masses in the continuum limit
(measured in units of the string tension
$\sqrt\sigma$) are listed in tab.1, where they are compared with the
corresponding results for SU(2) obtained in~\cite{tep} and with the predictions
of the IP model.

\begin{table}[ht]
\label{ip5f}
\caption{\sl Comparison between the  Ising, SU(2) and IP spectra.}
  \begin{center}
\begin{tabular}{|l|l|l|l|}
\hline
$J^P$ &  Ising & SU(2) & IP     \\
\hline
$0^+$  &  3.08(3)& 4.763(31) & 2.00  \\
$(0^+)'$ & 5.80(4)&   & 5.94  \\
$(0^+)''$  &  7.97(11)& & 8.35   \\
$2^{\pm}$  &  7.98(8) & 7.78(10)& 6.36  \\
$(2^\pm)'$  &  9.95(20)&  & 8.76  \\
$0^-$   &   10.0(5) & 9.90(27) & 13.82 \\
$(0^-)'$  & 13.8(6) &   & 15.05  \\
$(1/3)^\pm$   & 12.7(5)& 10.75(50) & 8.04  \\
\hline
\end{tabular}
  \end{center}
\end{table}
Looking at tab.1 we see that the biggest discrepancy in the mass values is for
the lowest state, which is predicted to be too light in the IP model, and turns
out to be very different in the Ising and SU(2) cases. This is due  first
to the lack of validity of the adiabatic approximation at small scales and
second to the fact that in the  IP model  an ``ideal'' picture of
string  (without self repulsion terms) is assumed  for the flux tube.

Apart from this state, in the remaining part of the spectrum 
we immediately see an impressive agreement between the Ising and SU(2) spectra.
This agreement is further improved by looking at the excited states in the  
$(0^+)$ channel for the SU(2) model. In~\cite{arisue} a variational estimate
 for these masses  can be found (up to our knowledge
no Montecarlo estimate exists for them). In tab.2 we compare these values with
the Ising ones. While the two sets of excited states disagree if measured in 
units of $\sqrt\sigma$, they agree if measured in units of $0^+$. 
Moreover a better and better agreement is observed if ratios of
higher masses are considered.

\begin{table}[ht]
\label{ip5g}
\caption{\sl The $0^+$ channel.}
  \begin{center}
\begin{tabular}{|l|l|l|}
\hline
ratio &  Ising & SU(2)      \\
\hline
$(0^+)'/0^+$ & 1.88(2)& 1.77(2)  \\
$(0^+)''/0^+$  & 2.59(4)& 2.50(5)    \\
$(0^+)''/(0^+)'$ & 1.37(4) &   1.41(4)   \\
\hline
\end{tabular}
  \end{center}
\end{table}
We can conclude from these data that the qualitative features of the 
glueball spectrum are largely independent from the gauge group and 
well described by a flux tube effective model.
While the higher states of the spectrum show a remarkable
independence from the gauge
group, for the lowest state the flux tube picture 
breaks down  and
the gauge group becomes important. The IP model, which is the simplest 
possible realization of such a flux
 tube,  seems  able to catch (at least at a qualitative level)
some of the relevant features
of the glueball spectrum.

\section{Duality.}
Another important reason of interest in the gauge Ising model is that it is 
related 
by duality to the ordinary $3D$ spin Ising model. As a consequence,
one expects the glueball spectrum to be mapped in the spectrum of 
massive excitations  of the spin
model. However this mapping has some non trivial
features. While the lowest state $0^+$ is
mapped into the (inverse) correlation length of the spin  model
(see~\cite{acch,ch} for details),
it is not clear which are the dual
partners of the higher states of the spectrum. In fact one naively expects that
the spin Ising model should be described in the scaling region by a $\phi^4$
theory, whose spectrum however contains just one state.
As a matter of fact the higher states of the gauge Ising model
correspond to ``disorder'' variables in the spin Ising model which are
{\sl non-local} with respect to the ``order'' variable (the $\phi$ field).
The overlap between these mutually non-local observables becomes
vanishingly small in the continuum limit (even if it is non-zero in any finite
lattice) thus explaining the difference between the two dual spectra. 
A detailed comparison of correlation functions in the Ising spin model and in 
the $\phi^4$
theory seems to confirm this picture~\cite{chp}.


\begin{thebibliography}{9}

\bibitem{ip} N.Isgur and J.Paton, \PR{D31} (1985) 2910.

\bibitem{acch} V.Agostini, G.Carlino, M.Caselle and M. Hasenbusch, 
\NP{B484} (1997) 331.

\bibitem{tep} M. Teper \PL{289B} (1992) 115,
T. Moretto and M. Teper, hep-lat/9312035

\bibitem{arisue}
H.Arisue, Prog. Theor. Phys. {\bf 84}  (1990) 951.


\bibitem{ch}  M.Caselle and M.Hasenbusch, 
J.Phys. {\bf A 30}  (1997) 4963.


\bibitem{chp}  M.Caselle, M.Hasenbusch and P.Provero, in preparation 


\end{thebibliography}
\end{document}